# GEANT4 simulation study of the response of a miniature radiation detector in Galactic Cosmic Rays and inside a spacecraft.

K. Karafasoulis[1,2], C. Papadimitropoulos[2], C. Potiriadis[3,2], C.P. Lambropoulos[2]

[1]Hellenic Army Academy, Vari, Greece

[2]Department of Aerospace Science and Technology, National and Kapodistrian University of Athens, Psahna-Evia, 34400 Greece

[3]Greek Atomic Energy Commission, Agia Paraskevi, Attiki, 15310 Greece

Abstract: The Miniaturized Detector for Application in Space (MIDAS) is a compact device with dimensions 5 x 5 x 1 $cm^3$ which combines position sensitive Si detectors and a fast neutrons spectrometer. MIDAS is developed with purpose to act as a linear energy transfer (LET) spectrometer for the charged particles and measure dose and dose equivalent from both charged particles and neutrons. It is based on fully depleted monolithic active Si pixel sensors for the charged track and energy deposition measurements, while a plastic scintillator read out by a silicon photomultiplier is used to determine energy depositions from fast neutrons. A simulation study of the detector response in galactic cosmic ray (GCR) radiation fields with the aid of GEANT4 has been performed. Energy depositions and hit pixel addresses have been used to reconstruct tracks and calculate LET spectra. A method to calculate $LET_\infty$ in water from the measured LET has been elaborated. Dose rate in water and dose equivalent rate have been calculated. The energy and particle composition of the radiation field produced by the interaction of GCR with the Al walls of a spacecraft model has been determined and the response of MIDAS in this radiation field has been investigated.

1. Introduction

Coarsely segmented Si diode detectors are used in almost all the instruments measuring the mixed radiation fields of the space environment (Spence H.E. et al.,2010; Hassler D.M. et al., 2012, Rodríguez-Pacheco, J. et al.,2020). For almost a decade, hybrid Si pixel detectors have started to penetrate the field of space dosimetry with starting event the installation of 5 Timepix units onboard the International Space Station in 2012 (Pinsky L, Hoang SM, 2014). Instruments using Si pixel detectors for monitoring radiation in the space environment can achieve orders of magnitude reduction in mass, size, and power consumption. Wide fields of view without compromising the geometric resolution can be achieved, something that cannot be done with coarsely segmented Si diodes in telescopic configurations. Miniaturized telescopes based on partially depleted active Si pixel sensors using the Silicon on Insulator technology with heritage from x-ray imaging have been proposed and are developed (Vrba et al., 2018).

It is well known that the absorbed dose of radiation is measured in units of Gray (1 Gy = 1 J/kg). As not all sources of radiation have the same biological effectiveness, the Dose Equivalent measured in units of Sieverts (Sv) takes this into account. The Dose Equivalent (in Sv) is equal to the Dose (in Gray) times the quality factor (Q), where Q is a function of the Linear Energy Transfer (LET) – the rate of energy loss of a particle in water measured in *keV·µm$^{-1}$*.

MIDAS is developed in response to the request of the European Space Agency (ESA) for a device whose size, power consumption, and radiation data output would increase the level of space-flight crew autonomy regarding operational decisions related to radiation hazards. ESA required a device sensitive to both charged particles and neutrons to enable measurement of dose, dose rate, energy deposition, LET, and calculation of dose equivalent. For neutrons, the desired energy ranges from 100 keV up to 200 MeV. For charged particles, the capability to measure LET values of at least *10 MeV·cm$^2$·mg$^{-1}$* in Si was also required. The mass limit of the device was set at 50 g, with dimensions up to 5 × 5 × 1 cm$^3$. Power consumption should allow 30 days of autonomous operation. For measuring the tracks and the energy depositions from charged particles we proposed to use in MIDAS fully depleted monolithic active pixel sensors (DMAPS). In the last decade, DMAPS have been studied and introduced in the high energy physics experiments for measuring minimum ionizing particles (Peric, 2007). Fully depleted Si sensors achieve fast and less prone to radiation damage charge collection. We designed in-pixel electronics to handle a wide dynamic range of energy depositions in Si voxels with dimensions 106 x 106 x 50 µm$^3$ from less than the most probable value of energy loss of minimum ionizing protons up to more than *2.33 MeV·µm$^{-1}$* (*10 MeV·cm$^2$·mg$^{-1}$* in Si). MIDAS is a cube whose 5 facets house 2 layers of DMAPS, while the one facet is covered by a silicon photomultiplier (SiPM) and one layer of DMAPS. The SiPM reads the light output of a plastic scintillator which occupies the interior volume of the cube with dimensions 7 x 7 x 7 mm$^3$. The five facets of the scintillator are covered by a Ti box with 1 mm thickness which absorbs the recoil protons with kinetic energy up to 18 MeV. The device concept, first prototype iteration, initial measurements and simulations have been described in (Lambropoulos et al., 2019). MIDAS is at technology readiness level 4.

In the present paper, based on simulation experiments performed with the aid of the GEANT4 (GEometry ANd Tracking) package, we present a detailed study of the following subjects: (a) The reconstruction of the charged particle tracks using the energy deposition patterns on the pixel layers. (b) The calculation of the measured LET distribution. (c) The correction of the measured LET for the escaped energy and the determination of $LET_\infty$ in water. (d) The calculation of the dose and dose equivalent rate using $LET_\infty$ in water. (e) The response of MIDAS inside a spacecraft model, where the GCR radiation field is modified due to the interaction with the Al walls. Although the purpose of the work presented is to elucidate some of the capabilities of this novel device with the aid of Monte Carlo tools, the obtained results agree with some field measurements and with simulation results reported by other groups.

2. The simulation environment, geometry, and materials

The Monte Carlo simulations were performed using the GEANT4 toolkit version 4.10.06.p02 (Agostinelli S. et al., 2003; Allison J. et al., 2006; Allison J. et al., 2016). The QGSP_BERT_HP physics list has been used. The acronym stands for the "Quark gluon string model", which allows to simulate reactions of high energy hadrons with nuclei and to simulate high energy electro- and gamma- nuclear reactions; the "Bertini cascade model", which treats nuclear reactions initiated by long-lived hadrons and gammas with energies between 0 and 10 GeV, and the "High precision neutron model", which uses cross-section for neutrons with energies up to 20 MeV from the ENDF/B-VI evaluated data library. It includes the G4EmStandardPhysics list and the production threshold for gammas, electrons and positrons is 1mm. For space applications the correct treatment of low-energy protons is important. For this reason, the default GEANT4 model (Lassila-Perini & Urbán, 1995) for the calculation of energy depositions by charged particles was checked: In a simulation experiment, protons with 1244.1 MeV kinetic energy were impinging vertically on a fully depleted silicon slab with 51 µm thickness. The parameters were selected purposely to compare with the values reported in Table VIII of (Bichsel, 1988) for protons with βγ=2.1 (momentum 2 GeV/c, kinetic energy 1244.1 MeV) and taken from (Bak, 1987). The peak of the distribution of the deposited energy was found at 13.82 keV and the Full Width at Half Maximum (FWHM) is 7.2 keV. The corresponding experimental values are 12.64 keV for the distribution peak of the deposited energy and 7.446 keV for the FWHM.

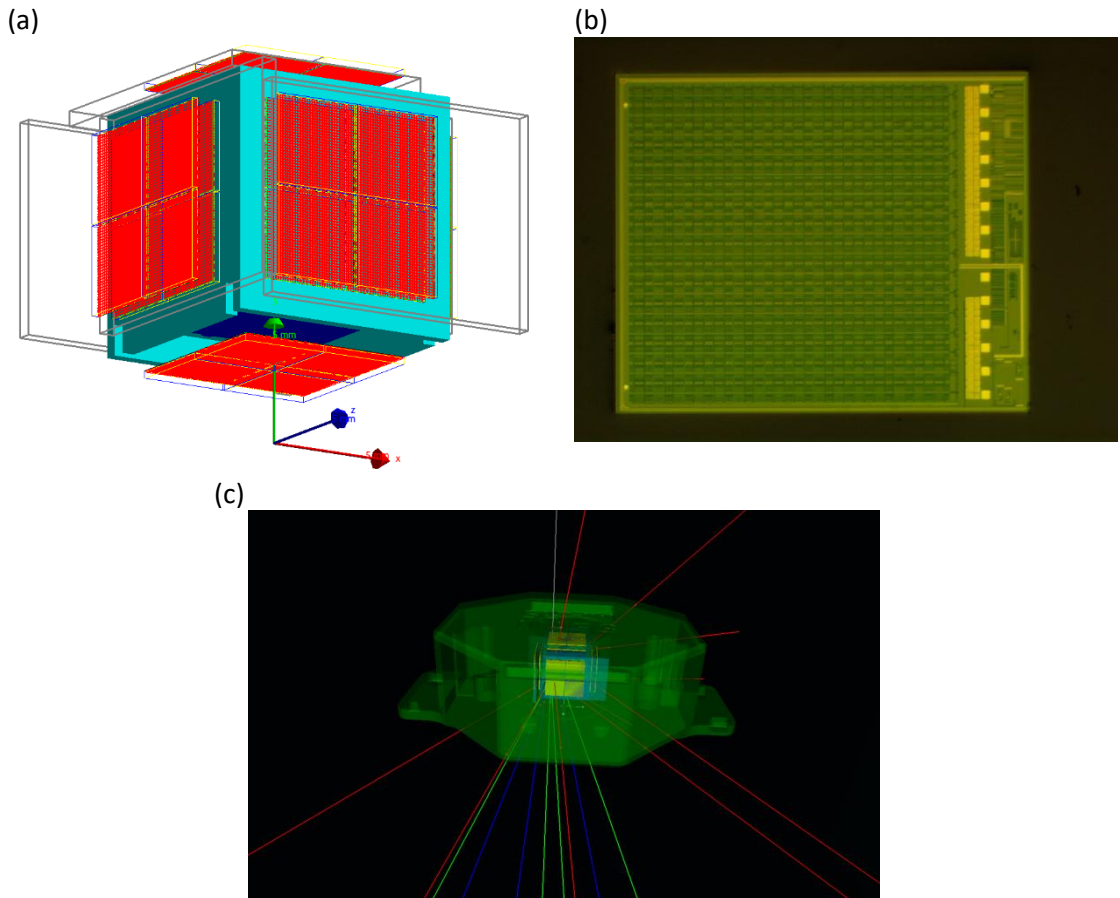

*Figure 1: (a) The geometry of the sensitive cube: Two layers of Si pixel detectors are placed on each facet of a Ti box, which contains a plastic scintillator. At the bottom a silicon photomultiplier in touch with the plastic scintillator and beneath them one layer of Si. (b) The die of the Si active pixel sensor. (c) Visualization of the tracks of secondary particles produced when a 10 GeV $^{56}$Fe ion hits the device. Gray: the $^{56}$Fe track; red: photon tracks; green: neutron tracks; blue: proton tracks*

With the aid of the OMERE freeware (OMERE), the differential flux distributions of galactic cosmic rays (GCR) according to the ISO 15390 model for the solar minimum 1996.4 between the end of Cycle 22 and the beginning of Cycle 23 were obtained and used as energy distributions of a spherical source with a radius of 20 cm. The energy range is from 0.1 MeV/n up to 20 GeV/n and it is partitioned in 835 logarithmic bins. The device was placed at the center of this source, which emitted particles with equal probability from all points of its interior surface, with cosθ and φ uniformly distributed in the interval [-1,1] and [0,2π] respectively, where θ is the polar angle with respect to the radius of the sphere from the point of emission and φ is the azimuthal angle in the plane perpendicular to this radius. Equal numbers of particles ($10^7$) were generated for each element with Z=1 up to Z=26. The distribution of a calculated quantity resulting from the Monte Carlo experiment for one element was scaled with the integral over energy of the element's flux with purpose to assign to this distribution the correct weight in the weighted sum for all particles. This weighted sum gives the final distribution of the quantity due to the GCR spectrum. The geometry of the device, the visualization of the tracks of the particles produced due to the passage of a Fe56 ion with 10 GeV kinetic energy and the die photo of the first version of the Si active pixel sensors can be seen in Figure 1.

The analysis presented here has been performed with the following assumptions: (a) The depleted monolithic active pixel sensors have 50 μm thickness, and all the energy deposited to a voxel with dimensions 106 x 106 x 50 μm³ is registered. The wafers of the final version of the MIDAS pixel detectors will be processed to become 50 μm thick. (b) A layer of silicon sensors is mounted below the SiPM and at the bottom layer of the associated electronics board. This means that all the plastic scintillator cube facets are covered by at least one silicon layer, something which was not foreseen in the initial design of the device, as explained in (Lambropoulos C.P. et al., 2019). (c) In the calculations only energy depositions above 20 keV are used, as a conservative estimate of the pixel readout electronics sensitivity.

3. Track reconstruction of charged particles.

The correct estimation of the track direction is necessary for the calculation of the distance traveled by the particle in a Si layer and, consequently, of the measured LET which can be defined as

$LET_{meas} = E_{dep}/L$ (1),

where $E_{dep}$ is the deposited energy in the Si sensor and $L$ is the track length in it. The deposited energy and the track length are quantities that can be found using either Monte Carlo or real data.

The track direction is estimated with the following steps:

(a) The pixels of a Si sensor with registered hits are grouped into clusters, provided they have a common edge or vertex, and the deposited energies in the voxels defined by theses pixels are summed. The cluster having the highest deposited energy is selected.

(b) As every Si layer contains 4 sensors, the cluster or isolated pixel with the highest deposited energy among all the 4 sensors is selected as a candidate for the 3D line fit that will provide the estimation of the track direction. For this cluster, the energy weighted barycenter is calculated.

(c) If energy has been deposited in more than four Si layers, there will be more than four candidate energy clusters. In this case, the barycenter points of the four most energetic ones are used as candidate points to which a 3-d line is fitted.

(d) Since heavy ions produce secondary particles in interactions with either the active (silicon and plastic scintillator) or the non-active materials (titanium box, enclosure, substrate PCBs) of the device, it is possible that the Si sensors register hits due to the secondary radiation and not from the primary track. Usually, the energy depositions due to delta rays are much lower than the ones produced by the primary ion. We found that by excluding from the 3D line fit clusters with energy less than 7% of the total energy of the four candidates, the distribution of the difference angle between the real and the reconstructed direction of the primary track becomes narrower.

(e) A 3D line fit is performed to the barycenter points of the clusters that have passed the cut imposed in step (d), provided they are more than two. If two clusters have remained, then the line that connects their barycenter points is calculated.

(f) When a 3D line fit is performed, the sum of the squares of the residuals of the fit is computed. If the sum is above 0.01 mm$^2$ the least energetic cluster is excluded, and the line is estimated again using the remaining clusters.

(g) The direction cosines of a unit vector along the estimated track line are calculated.

Using the described algorithm, a track direction is calculated for every event and this is considered as the direction to be used for the determination of LET. However, it is probable that a primary ion will interact with the non-active parts of the device and the Si sensors will register only the secondary tracks and not the primary one. As in reality it will not be possible to discriminate against such an event, we do not exclude the false primary tracks from our analysis. This practically means that a highly energetic delta ray may be treated as if it were a proton. In Figure 2 is presented the distribution of the angle between the real and the reconstructed track for a sample of $^{56}$Fe events with the GCR energy distribution. One can see that although the distribution has peaks at 0 rad and at π rad, there is a continuum of events between these two angles, because the reconstructed track assigned to a primary particle is the track of a secondary delta ray.

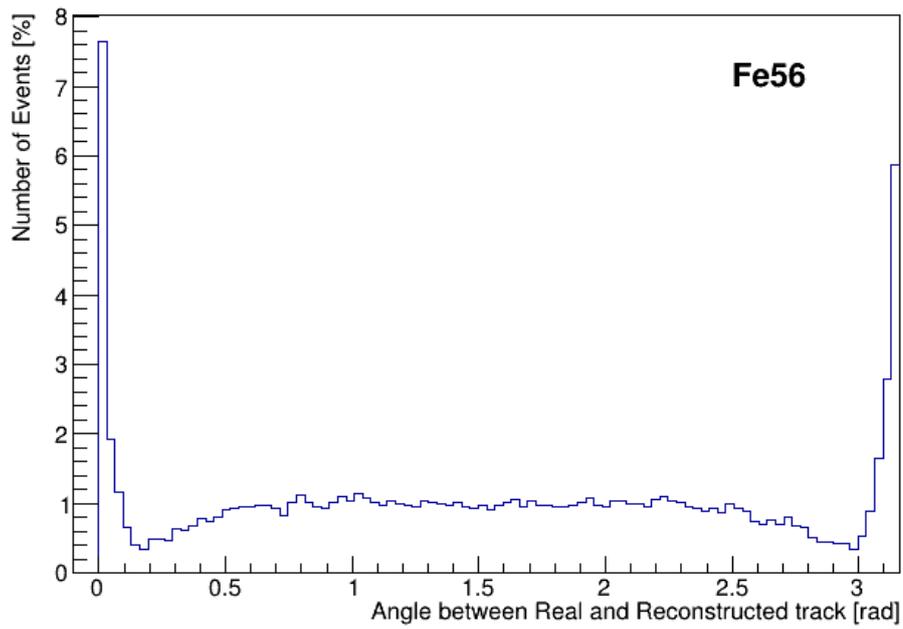

*Figure 2: Distribution of the angle between the real (known from MC) and reconstructed track*

4. LET estimation

The measured LET is calculated using (1) for each Si layer traversed by the fitted track line and the average value for these layers is used to construct the distribution of $\text{LET}_{\text{meas}}$. As this distribution has been calculated separately for each sample of primary ions, it is scaled with the ion's relative contribution to the GCR flux spectrum and the weighted sum for all of them is the final measured $LET_{meas}(Si)$. The result is presented in Figure 3 together with $LET_{H_2O}$ deduced from $LET_{meas}$ with the procedure described below.

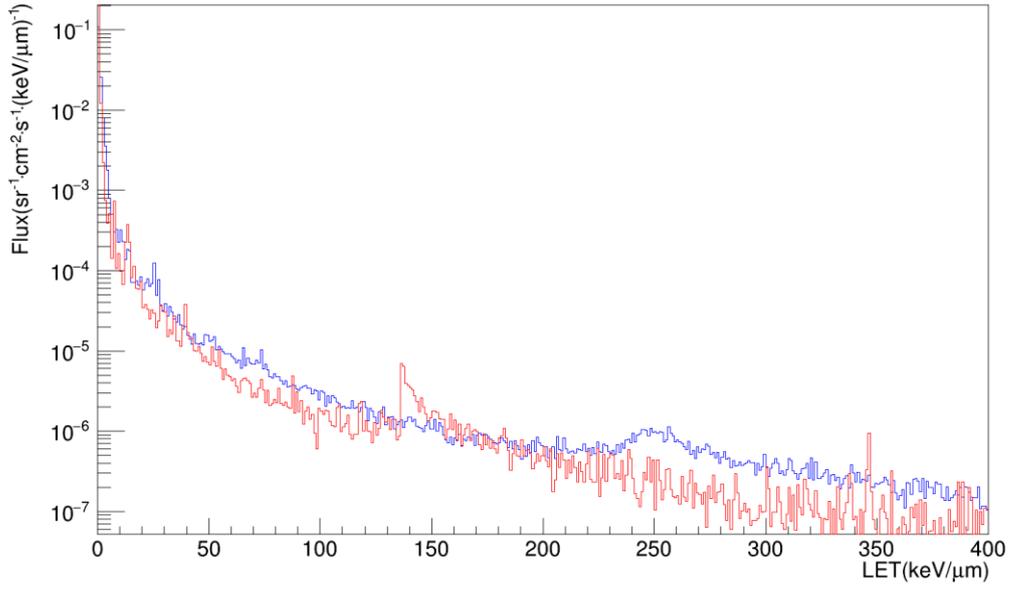

*Figure 3: $LET_{meas}(Si)$ resulting from the addition of the weighted distributions from all the ions with Z=1 to Z=26 using as weights their relative contribution to the GCR flux spectrum (blue) and $LET_{H_2O}$ computed using the method described in the text. The bin width is 1 keV·µm$^{-1}$*

For each sample of ion species, the Monte Carlo truth information is used to calculate the LET in water $(LET_{H_2O})$ using the Geant4 G4EMCalculator for every reconstructed track.

$LET_{meas}$ and $LET_{H_2O}$ for all ion species are inserted in the scatter plot presented in Figure 4. The best fit line to this diagram is found to be

$$log_{10}(LET_{H_2O}) = p_0 + p_1 \cdot log_{10}(LET_{meas}) \quad (2)$$

Where $p_0 = -0.25797 \pm (0.00085)$ and $p_1 = 0.98514 \pm (0.00038)$. With the aid of this equation the obtained $LET_{meas}$ values are converted to $LET_{H_2O}$ values. This result is very close to that by Benton et al. (2010) obtained from range/energy relations for ions ranging in charge from 1 to 26 over an energy interval of 0.8-2000 MeV·amu$^{-1}$

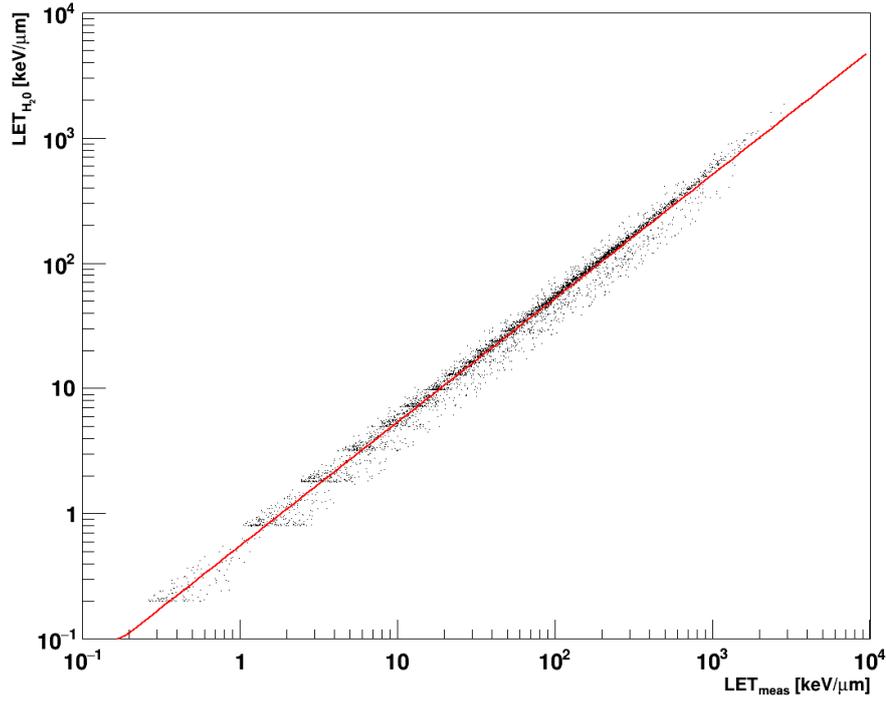

Figure 4: Scatter plot of $LET_{H_2O}$ as calculated with G4EMCalculator and of $LET_{meas}$ for all ions. A reduced number of points for each sample is included to improve visibility. The fitted line is computed using the full samples.

The measured LET spectrum should be corrected both for inefficiencies in the estimation of the primary tracks and for the wrong inclusion of secondary or, sometimes, fake tracks as primaries. The reconstruction efficiency can be defined as the number of simulated events whose reconstructed tracks contribute to the measured LET spectra over the events handled by the Geant4 G4EMCalculator. Consequently, correction factors $C_{j,rec}$ as a function of $LET_{H_2O}$ are calculated as the inverse of the reconstruction efficiency. They are given by the following expression:

$$C_{j,rec}(LET_{H_2O}) = \frac{\sum_{ions} w_{ion} \cdot \hat{n}_{j,ion}(LET_{H_2O})}{\sum_{ions} w_{ion} \cdot n_{j,ion}(LET_{H_2O})}$$

where the sum is over all the ions contributing to the simulation, $w_{ion}$ is the relative contribution of the ion in the GCR spectrum, $\hat{n}_{j,ion}(LET_{H_2O})$ is the content of the bin j of the LET distribution constructed using the Geant4 G4EMCalculator to find the values of LET in water from all the primary particles entering the silicon volumes, while $n_{j,ion}(LET_{H_2O})$ is the content of bin j of the measured LET distribution, calculated from the reconstructed tracks and converted using equation (2). $C_{j,rec}(LET_{H_2O})$ have been calculated from samples of events different from those used to extract the measured LET distribution to which they have been applied.

Consequently, the $LET_{H_2O}$ distribution presented in Figure 3 and the dose estimation described in the following section have been deduced using the content $n_j(LET_{meas})$ of each bin j of the measured LET distribution (either from a real or a Monte Carlo sample) converted using equation (2) to $n_j(LET_{H_2O})$ and multiplied with $C_{j,rec}(LET_{H_2O})$.

5. Dose and dose equivalent calculation

As the purpose of this study is to provide estimates of the dose and dose rate indications of the MIDAS device in the GCR field, a computation of count rate is necessary. For this reason, the effective cross section of the device has been determined with the following procedure: In the interior of the spherical source used to generate $10^7$ primary particles with their directions distributed uniformly, a series of concentric spheres were placed. In these spheres the primary particles do not undergo any interaction; they are used only for counting the primary particles entering their interior. The area of the great circles of the concentric spheres as a function of the number of the particles entering them shows a linear behavior, thus a linear equation relating the area of the great circle to the number of entering particles was obtained. The number of particles entering the MIDAS sensitive volume without surrounding enclosure was counted with the aid of a separate Monte Carlo experiment. It was found that the number of particles entering MIDAS is the same with those entering a sphere with *0.422 cm* radius and great circle with *0.5595 cm²* area, which is the effective cross section, $A_{eff}$, of MIDAS. The time corresponding to the $10^7$ particles generated was deduced using the relation

$$time = \frac{N_{incident}}{A_{eff} \cdot \Phi_{tot}} \quad (3),$$

where $N_{incident}$ is the number of particles entering MIDAS and $\Phi_{tot}$ is the total integral flux of all ions with Z=1 up to Z=26 provided by the OMERE freeware.

The $LET_{H_2O}$ spectrum is then normalized in order to be expressed as $particles \cdot cm^{-2} \cdot s^{-1}$ and the dose rate is calculated as

$$D = \frac{1}{\rho_{H2O}} \sum_{i=1}^{N} LET_i \cdot N_i \quad (4),$$

where $N_i$ is the content of bin i of the $LET_{H_2O}$ spectrum expressed in $particles \cdot cm^{-2} \cdot s^{-1}$, $LET_i$ is the LET at the center of bin i expressed in $joules \cdot cm^{-1}$ and $\rho_{H_2O}$ is the water density in $kg \cdot cm^{-3}$. The LET bin width has been set to *1 keV·µm⁻¹* and the sum was performed for LET values up to *400 keV·µm⁻¹* and for LET values up to *1000 keV·µm⁻¹*.

We have found an absorbed dose rate in water $D = 0.506 \, mGy \cdot day^{-1}$ when summing up to *400 keV·µm⁻¹* and $D = 0.526 \, mGy \cdot day^{-1}$ when summing up to *1000 keV·µm⁻¹* for the GCR spectrum.

For the dose equivalent calculation, we have used the (ICRP, 1991) quality factors Q:

$$Q(L) = \begin{cases} 1 & L < 10 keV \cdot \mu m^{-1} \\ 0.32L - 2.210 keV \cdot \mu m^{-1} & 10 keV \cdot \mu m^{-1} \leq L \leq 100 keV \cdot \mu m^{-1} \\ 300/\sqrt{L} & L > 100 \mu m^{-1} \end{cases} \quad (5),$$

and the relation

$$H = \frac{1}{\rho_{H2O}} \sum_{i=1}^{N} LET_i \cdot Q(LET_i) \cdot N_i \quad (6)$$

We have found a dose equivalent rate in water for the GCR spectrum $H = 2.32 \, mSv \cdot day^{-1}$ when summing up to 400 *keV·µm⁻¹* with an average Quality factor *<Q>= 4.58* and $H = 2.57 \, mSv \cdot day^{-1}$ with *<Q>= 4.89*, when summing up to *1000 keV·µm⁻¹*.

In the correction calculation of the inefficiency of MIDAS in reconstructing the LET spectrum the model of the radiation environment is used. As in reality the radiation environment is not known beforehand, we tried to estimate the effect on the dose calculation using the measured LET distribution for the 1996.4 solar minimum corrected with $C_{j,rec}$ evaluated for solar maximum. The resulting dose rate in water is *0.460397 mGy·day⁻¹*, the dose equivalent rate in water is *1.99588 mSv·day⁻¹* and *<Q> = 4.33513*

6. Radiation field inside the spacecraft

The radiation field inside a spacecraft is modified due to the partial absorption of the primary particles coming from GCR and SEP (solar energetic particles) events in the walls and the reactions of these particles with the spacecraft materials. Therefore, the response of MIDAS in a radiation field representative of what will be encountered inside a spacecraft has been studied.

The field distributions were constructed using a spacecraft model placed inside a spherical source with radius of 1 m. The spacecraft was bombarded with particles having the same differential flux spectrum as in the calculations presented in the previous sections. The simplified spacecraft model is a cylinder whose both inner radius and height are 80 cm. The material of its walls consists of 7.407 cm (20 g/cm$^2$) thick Al. Its interior is filled with air. A probing sphere with 20 cm radius is placed at the cylinder's center to record the incoming particles and their energies. The particles considered for the construction of the internal field distributions to be used are nuclei with Z=1 to Z=26 plus 12 non-existing in the primary field species produced from the interactions of the GCR with Al and air. The additional 12 particles are neutrons, gammas, $e^-/e^+$, $\mu^-/\mu^+$, $K^-/K^+$, $\pi^-/\pi^+$, deuterons and tritons. For each particle its relative contribution to the internal field was calculated. In Figure 5 (a) are shown the ratios of the flux relative contributions of the charged particles present in both the GCR and the spacecraft field. The error bars represent statistical fluctuations of the Monte Carlo samples used. In Figure 5 (b),(c) are presented the relative contribution in the flux and in the dose in water respectively of all the particles of the spacecraft field. Dose in water has been calculated with the aid of the G4EMCalculator. Figure 5(d) shows energy spectra of the 4 particles which are not present in the GCR external field, but they have the highest contribution in flux.

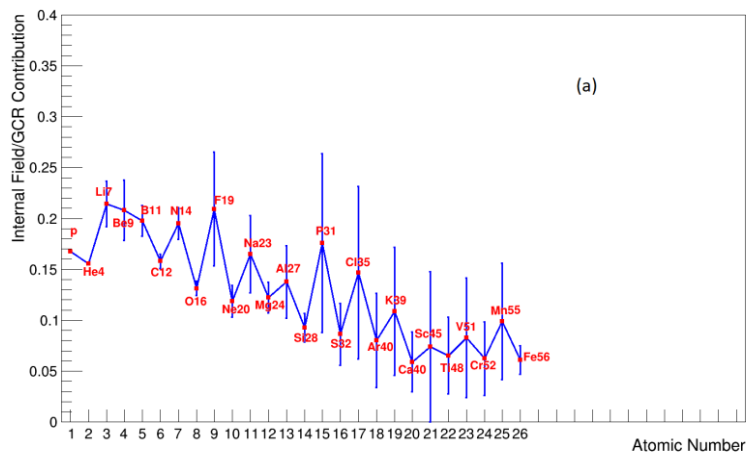

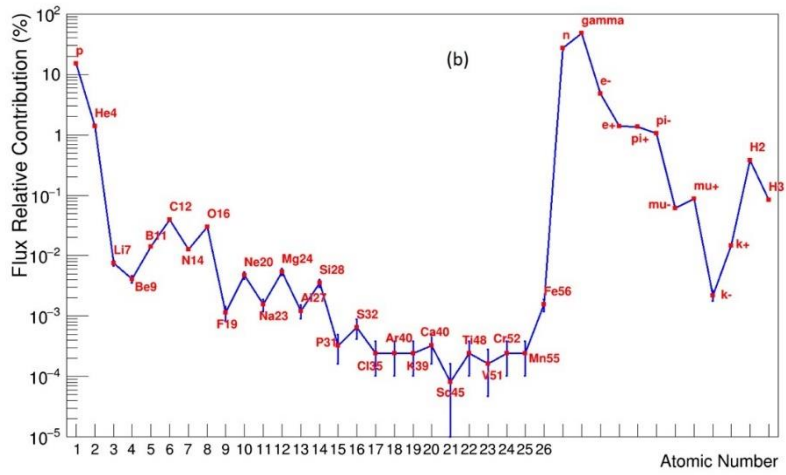
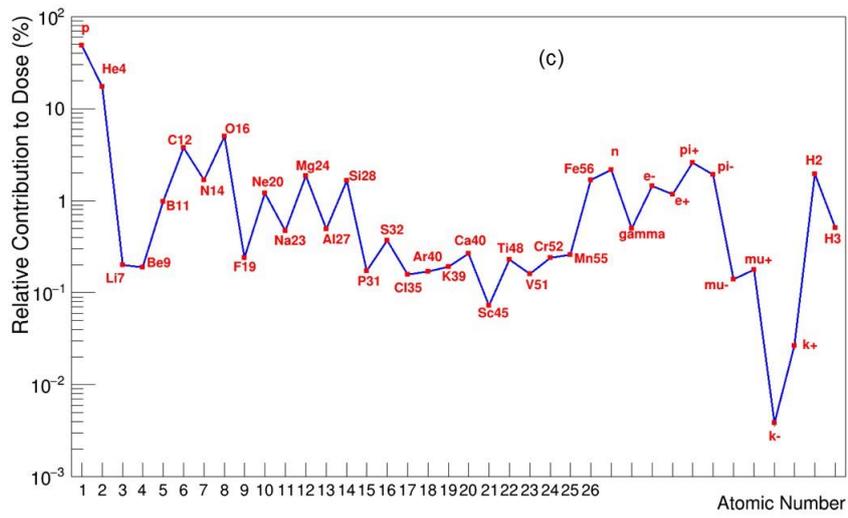
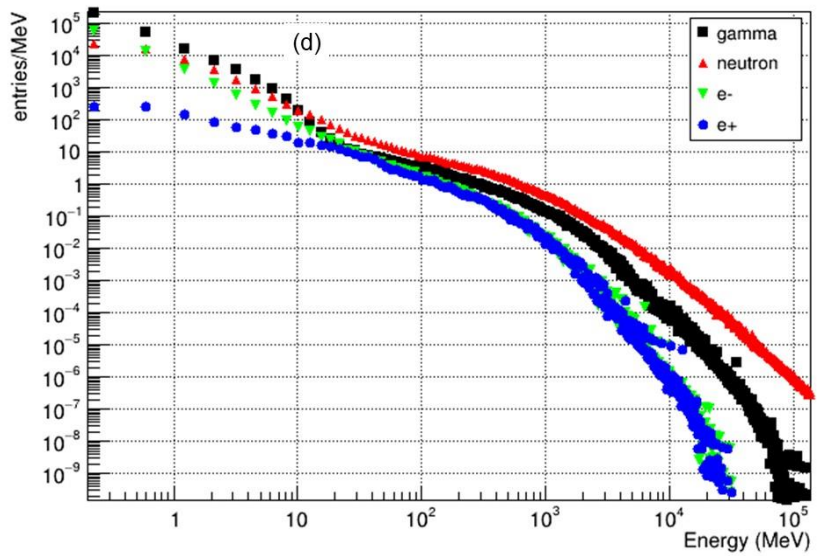

*Figure 5. (a) For the particles with Z=1 to Z=26 the ratio of their contribution in the integrated flux of the spacecraft internal field over their contribution in the integrated flux of the GCR field is shown. (b) The relative contribution of the particles in the integrated flux of the spacecraft internal radiation field is presented. (c) The relative contribution to dose in water of the particles of the spacecraft internal radiation field. (d) Energy spectra for gammas, neutrons, electrons, and positrons, which are the most abundant species in the internal field.*

The resulting energy spectra were used as inputs for the definition of the spherical sources surrounding MIDAS. Equal numbers of particles ($10^7$) were generated for each species. Again, the distribution of a calculated quantity resulting from the Monte Carlo experiment for one particle species was scaled with its relative contribution and the sum for all species was derived.

7. Estimation of LET and dose inside the spacecraft

The measured LET was estimated using (1) and converted to LET in water with the procedure described in the previous paragraphs. It is shown in Figure 6. With purpose to check that equation (2) can be used for the internal field too, the fit was repeated for the corresponding $LET_{meas}$ and $LET_{H_2O}$ and the result was almost identical. The count rate calculation for the internal field has been performed. For each GCR particle species $25 \cdot 10^6$ events were generated and the number of particles either primary or secondary entering the probing sphere was counted, weighted with the relative contribution in the GCR flux of the parent species and summed. The ratio of this number to $25 \cdot 10^6$ is the estimate of the coefficient which relates any number of particles generated from a spherical source with radius of 20 cm and with the composition and energy distribution of the internal field to the number of the primary GCR particles sample from which it originates. This coefficient is used to retrieve an effective $N_{incident}$ in (3).

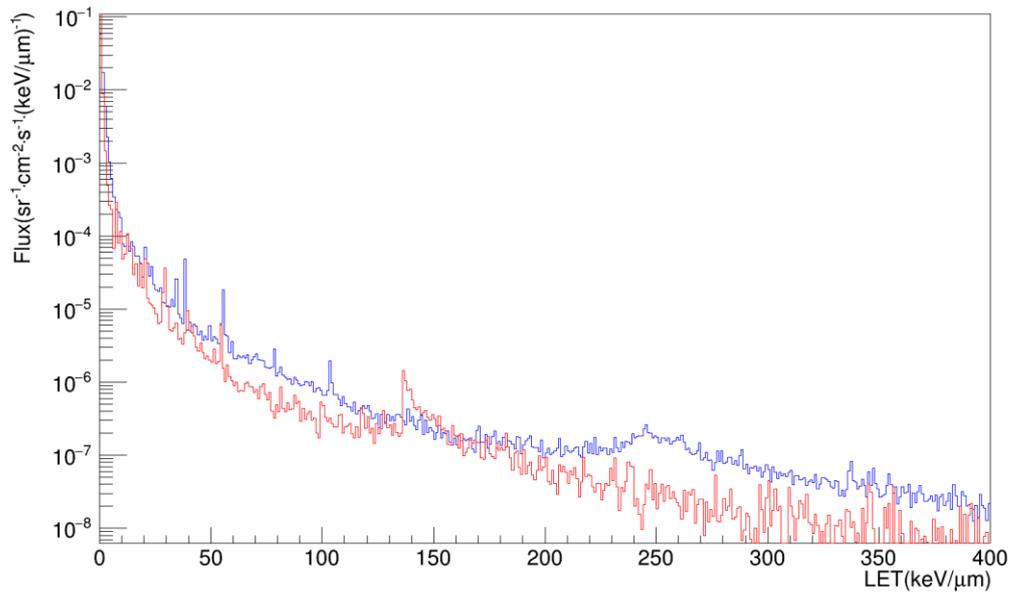

*Figure 6. $LET_{H_2O}$ spectrum for the field inside the spacecraft (red) and $LET_{exp}(Si)$ (blue). Both are calculated using the energy depositions in the pixel detectors of MIDAS*

The resulting absorbed dose rate in water is *D= 0.25 mGy·day$^{-1}$*, for integration up to *1000 keV·µm$^{-1}$* and *D=0.2486 mGy·day$^{-1}$* for integration up to *400 keV·µm$^{-1}$*. The corresponding dose equivalent rates and quality factors are *H=0.654 mSv·day$^{-1}$*, *<Q> =2.61* for integration up to *1000 keV·µm$^{-1}$* and *H=0.634 mSv·day$^{-1}$*, *<Q>=2.55* for integration up to 400 keV·µm$^{-1}$.

## 8. Discussion

The work presented concerns the development of the methodology to extract LET spectra, dose, and dose equivalent from the energy depositions by charged particles in a novel device based on Si pixel detectors. The radiation fields used in the simulation represent what could be encountered in flight outside the geomagnetic field without shielding and inside a spacecraft. Obviously, the conditions in a real flight will vary, but the qualitative characteristics of the fields will not be different from those used in this study. The particles and their energies encountered in a solar energetic particle (SEP) event are a subset of those included in the study. In the case of SEP events, the high rate may affect the response of an instrument. Pixel detectors are advantageous in coping with high rates because of the independent signal processing within each pixel.

The end results obtained from this analysis of the simulated response of MIDAS are in fair agreement with measurements and simulations reported for instruments which have been flown and have an order of magnitude higher dimensions mass and power consumption. This fact provides evidence that the methodology presented is valid and on the other hand it accentuates the potential of this miniaturized LET spectrometer.

The dose equivalent rate on the Lunar surface reported by the LND experiment (Zhang et al., 2020) for charged particles is *1.37 mSv·day$^{-1}$* and *<Q> is 4.3*. Dose equivalent due to GCR charged particles on the lunar surface must be multiplied by 2, when compared to the dose equivalent in interplanetary space. The quality factor found by us for the field inside the spacecraft is almost the same with the quality factor for the same shielding thickness (20 g·cm$^{-2}$) calculated using the PHITS Monte Carlo code and presented in Figures B1 and B2 of (Zeitlin C. et al., 2019). Fig. 7 of (Banjac et al., 2019) reports comparable values for dose rates in a water slab for the ϕ=420 MV parameter of the BON model used to model the GCR flux of the 1996.4 solar minimum.

In Table 1 are collected the results of the calculations presented in this paper and the results from measurements and simulations with which they are compared.

*Table 1: Dose rates and quality factors from the MIDAS simulated response in 1996.4 solar minimum calculated using correction coefficients resulting from solar minimum and from solar maximum GCR fields. They are compared with the measurements on the lunar surface which are multiplied by 2. The 3rd column reports simulation results obtained using the Badwar O'Neil model and completely different methodologies than the one reported.*

| | MIDAS simulation result for GCR field at 1996.4 Solar minimum | Measurement on the Lunar surface (Zhang et al., 2020) | |
|---|---|---|---|
| Dose rate in water (LET limit $400$ keV·$\mu m^{-1}$) | *0.460 – 0.506 mGy·day$^{-1}$* | *0.318 ± 0.034 mGy·day$^{-1}$* <br> x 2 <br> (0.636 ± 0.068) | Simulation result by Banjac et al. 2019: <br> *0.6 mGy·day$^{-1}$* |
| Dose equivalent rate in water (LET limit $400$ keV·$\mu m^{-1}$) | *2.32 – 1.99 mSv·day$^{-1}$* | *1.37 ± 0.25 mSv·day$^{-1}$* <br> x 2 <br> (2.74 ± 0.5) | |
| <Q> (LET limit $400$ keV·$\mu m^{-1}$) | *4.58 - 4.33* | *4.3 ± 0.7* | |
| <Q> under 20 g·cm$^{-2}$ of Al shield | *2.55* (LET limit $400$ keV·$\mu m^{-1}$) <br> *2.61* (LET limit $1000$ keV·$\mu m^{-1}$) | | Simulation result by Zeitlin et al., 2019: 2.3-2.4 |

The contribution of neutrons or of other particles present in the radiation field inside a spacecraft have not been considered in the calculations as well as in the results of other works which have been used for the comparisons.

The calculations have been based on clusters of energy depositions in Si voxels with lateral dimensions equal to the pixel pitch and 50 µm thickness. This assumption considers the spread of energy deposition due to the delta rays, but it does not account for the pixel detector non idealities such as the charge diffusion and the fluctuations of gain or other systematic and random effects. These effects can be included in the simulation when a working prototype of the Silicon pixels detectors will be experimentally characterized.

## Acknowledgment

This work has been funded by the European Space Agency Contract 4000119598/17/NL/LF for the development of a highly miniaturized ASIC radiation detector. Also, it has been supported by computational time granted from the Greek National Infrastructures for Research and Technology S.A. (GRNET) in the National HPC facility – ARIS.

## References

Agostinelli S., J. Allison J., Amako K., Apostolakis J., Araujo H., et al. 2003. GEANT4—A simulation toolkit, Nucl. Instrum. Methods Phys. Res. Sect. A, 506(3), 250–303, https://doi.org/10.1016/S0168-9002(03)01368-8.

Allison J., Amako K., Apostolakis J., Araujo H., Arce Dubois P., et al. 2006. Geant4 developments and applications. IEEE Trans. Nucl. Sci., 53 (2006), p. 270. https://doi.org/10.1109/TNS.2006.869826

Allison J., Amako K., Apostolakis J., Arce P., Asai M., Aso T., et al.2016. Recent developments in GEANT4. Nucl Instrum Meth A 835: 186–225. https://doi.org/10.1016/j.nima.2016.06.125.

Bak J.F., Burenkov A., Petersen J.B.B., Uggerhoj E., Moller S.P., Siffert P. 1987. Large Departures from Landau Distributions for High Energy Particles Traversing Thin Si and Ge Targets. Nucl. Phys B288 681. https://doi.org/10.1016/0550-3213(87)90234-3

Banjac S., Berger L., Burmeister S., Guo J., Heber B., et al. 2019. Galactic Cosmic Ray induced absorbed dose rate in deep space – Accounting for detector size, shape, material, as well as for the solar modulation. J. Space Weather Space Clim. 9, A14. https://doi.org/10.1051/swsc/2019014

Benton E. R., Benton E. V., Frank A. L. 2010. Conversion between different forms of LET. Radiat Meas 45(8): 957–959. https://doi.org/10.1016/j.radmeas.2010.05.008

Bichsel H. 1988. Straggling in thin silicon detectors. Reviews of Modern Physics, 60(3) (1988). https://doi.org/10.1103/RevModPhys.60.663

Geant4-Collaboration. 2016. Physics reference manual – version: Geant4 10.3 (9 December 2016). http://geant4-userdoc.web.cern.ch/geant4-userdoc/UsersGuides/PhysicsReferenceManual/fo/PhysicsReferenceManual.pdf.

D.M. Hassler D. M., Zeitlin C., Wimmer-Schweingruber R.F., Böttcher S., Martin C., et al. 2012. The Radiation Assessment Detector (RAD) Investigation. Space Sci Rev 170:503–558. https://doi.org/10.1007/s11214-012-9913-1

ICRP, 1991. 1990 Recommendations of the International Commission on Radiological Protection. ICRP, Publication 60. Ann. ICRP 21(1–3)

ICRP, 2013. Dietze, G., Bartlett, D.T., Cool, D.A., Cucinotta, F.A., Jia, X., McAulay, I.R., Pelliccioni, M., Petrov, V., Reitz, G., Sato, T. 2013. ICRP PUBLICATION 123: Assessment of radiation exposure of astronauts in space. Annals of the ICRP 42 (4), 1–339. https://doi.org/10.1016/j.icrp.2013.05.004.

Lambropoulos C.P., Potiriadis C., Theodoratos G., Kazas I., Papadimitropoulos C., et al. 2019. MIDAS: A Miniature Device for Real-Time Determination of the Identity and Energy of Particles in Space. Space Weather, 18, e2019SW002344. https://doi.org/10.1029/2019SW002344

Lassila-Perini, K., Urbán, L. 1995. Energy loss in thin layers in GEANT. Nucl. Instrum. Methods Phys. Res. Sect. A, 362, 416–422. https://doi.org/10.1016/

Peric I. 2007. A novel monolithic pixelated particle detector implemented in high-voltage CMOS technology. Nuclear Instruments and Methods in Physics Research A 582 876–885. https://doi.org/10.1016/j.nima.2007.07.115

Pinsky L., Hoang S. M., Idarraga-Munoz J., Kruppa M., Stoffle N., et. al. 2014. Summary of the first year of medipix-based space radiation monitors on the ISS. 2014 IEEE Aerospace Conference, pp. 1-8, https://doi.org/10.1109/AERO.2014.6836502

Rodríguez-Pacheco J., Wimmer-Schweingruber R. F., Mason G. M., Ho G. C., Sánchez-Prieto S., et al. 2020. The Energetic Particle Detector. Energetic particle instrument suite for the Solar Orbiter mission. Astronomy and Astrophysics, 642, A7. https://doi.org/10.1051/0004-6361/201935287

Spence H.E., Case A. W., Golightly M. J., Heine T., Larsen B.A., et al. 2010. CRaTER: The Cosmic Ray Telescope for the Effects of Radiation Experiment on the Lunar Reconnaissance Orbiter Mission. Space Sci Rev 150: 243–284. https://doi.org/10.1007/s11214-009-9584-8

Zhang S., Wimmer-Schweingruber R. F., Yu J., Wang C., Zou Y., et al. 2020. First measurements of the radiation dose on the lunar surface. Sci. Adv. Vol. 6 no. 39. https://doi.org/10.1126/sciadv.aaz1334

Zeitlin C., Narici L., Rios R. R., Rizzo A., Stoffle N., et al. 2019. Comparisons of High-Linear Energy Transfer Spectra on the ISS and in Deep Space. Space Weather, 18, e2019SW002344. https://doi.org/10.1029/2018SW002103

Vrba V., Benka T., Fojtik J., Havranek M., Janoska Z., et al. 2018. The SpacePix-D radiation monitor technology demonstrator. JINST 13 C12017. https://doi.org/10.1088/1748-0221/13/12/C12017